\documentclass{elsart5p}

\usepackage{graphics}
\usepackage{graphicx}
\usepackage{amssymb}
\begin{document}

\begin{frontmatter}



\title{Elastic neutron scattering in Quantum Critical Antiferromagnet Cr$_{0.963}$V$_{0.037}$.}
%

\author[AA]{D. A. Sokolov\corauthref{D. A. Sokolov}},
\ead{sokolov@bnl.gov}
\author[AA,BB]{M. C. Aronson},
\author[CC]{G. L. Strycker},
\author[DD]{M. D. Lumsden},
\author[DD]{S. E. Nagler},
\author[EE]{R. Erwin}

\address[AA]{Brookhaven National Lab. Upton, NY 11973, USA}
\address[BB]{Department of Physics and Astronomy, State University of New York, Stony Brook, New York 11794, USA}
\address[CC]{Department of Physics, University of Michigan, 450 Church Street, Ann Arbor, MI 48109-1040, USA}
\address[DD]{HFIR, Oak Ridge National Lab., TN 37831, USA}
\address[EE]{NIST Center for Neutron Research, NIST, Gaithersburg, MD 20899, USA}

\corauth[D. A. Sokolov]{Corresponding author. Tel: (631) 344-2844
fax: (631) 344-4071}

\begin{abstract}
We have performed elastic neutron scattering studies of the quantum
critical antiferromagnet Cr$_{0.963}$V$_{0.037}$.  We have found
that unlike pure Cr, which orders at two incommensurate wavevectors,
Cr$_{0.963}$V$_{0.037}$ orders at four incommensurate and one
commensurate wavevectors. We have found strong temperature dependent
scattering at the commensurate and incommensurate wavevectors below
250 K. Results indicate that the primary effect of V doping on Cr is
the modification of the nesting conditions of the Fermi surface and
not the decreasing of the Neel temperature.
\end{abstract}


\end{frontmatter}


Pure Cr is a body centered cubic itinerant antiferromagnet, which
orders at 311 K via spin-density wave instability. The electron and
hole octahedra, which compose the Fermi surface of Cr are nested by
the q=(1$\pm$$\delta$00)2$\pi$/$\emph{a}$ wavevectors, where
$\emph{a}$ is a lattice constant of Cr~\cite{fawcett1988}. In the
elastic neutron scattering experiment the antiferromagnetism in Cr
is marked by resolution limited superlattice peaks at the nesting
wavevectors q=(1$\pm$$\delta$00)2$\pi$/$\emph{a}$, which appear at
the Neel temperature ($\emph{T}$$_{N}$). It has been reported that
doping Cr with V reduces $\emph{T}$$_{N}$, and at V concentrations
close to 4$\%$ $\emph{T}$$_{N}$$\rightarrow$0~\cite{takeuchi1980}.
The Neel temperature in CrV alloys is usually determined by the
minimum in the temperature dependence of the electrical resistivity,
usually attributed to removal of the parts of the Fermi surface due
to nesting. The possibility of generating a quantum critical point
(QCP) in a simple metal motivated a number of experimental and
theoretical studies of quantum criticality in
Cr~\cite{yeh2002,lee2004,norman2003}. The main evidence for a QCP in
CrV in addition to $\emph{T}$$_{N}$$\rightarrow$0, is the increase
of the number of carriers when doping into the paramagnetic state,
determined from the Hall effect measurements~\cite{yeh2002}. In this
work we report the results of the first neutron scattering
measurements performed on a high quality single crystal of nominally
quantum critical Cr$_{0.963}$V$_{0.037}$.

Single crystals of Cr$_{1-x}$V$_{x}$, x=0.0, 0.02, 0.037 were grown
by the arc zone melting method at the Materials Preparation Center
at Ames National Lab. Electron microscopy measurements confirmed
that the V concentration remains uniform on length scales from 1 mm
to 50 nm. Electrical resistivity $\rho$ of Cr$_{1-x}$V$_{x}$, x=0.0,
0.02, 0.037 was measured by a conventional four probe method using a
PPMS by Quantum Design. Neutron scattering experiments were carried
out at the NIST Center for Neutron Research on 40 g single crystal
of Cr$_{0.963}$V$_{0.037}$ and 30 g single crystal of pure Cr using
BT9 triple-axis spectrometer with a fixed incident energy
E$_{i}$=14.7 meV. The measurements were done using a
40'-44'-44'-open collimation configuration. Data were collected near
the (100) reciprocal lattice position in the (011) plane. Some of
our results were obtained on the HB3 triple-axis spectrometer at
HFIR, ORNL.

Fig.~1 shows the temperature dependence of the electrical
resistivity $\rho$ of single crystals of Cr$_{1-x}$V$_{x}$, x=0.0,
0.02, 0.037. $\rho$ for x=0.0 and x=0.02 were normalized to yield
the same value as $\rho$ for x=0.037 at 340 K. Cr$_{1-x}$V$_{x}$
alloys are good metals with residual resistivity $<$1
$\mu$$\Omega$$\cdot$cm. The Neel temperature ($\emph{T}$$_{N}$) is
marked by a minimum in $\rho$(T), which becomes broader and shifts
to lower temperature as the V concentration increases. The inset in
Fig.~1 shows $\emph{T}$$_{N}$ of Cr$_{1-x}$V$_{x}$ as a function of
V concentration of single crystals reported in this work and in
previously reported studies of polycrystalline
Cr$_{1-x}$V$_{x}$~\cite{takeuchi1980}. The results of this work
and~\cite{takeuchi1980} are in a good agreement. Results of
transport measurements indicate that doping Cr with V reduces
$\emph{T}$$_{N}$ and make the antiferromagnetism disappear at
x$\geq$0.037.

\begin{figure}
\includegraphics[scale=0.42]{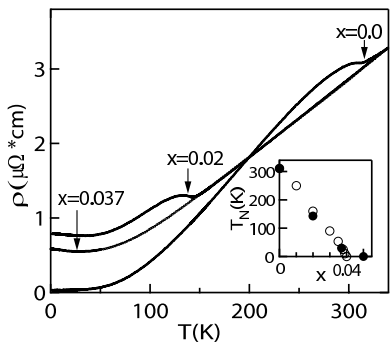}
\caption{\label{fig:epsart} The temperature dependence of the
electrical resistivity of Cr$_{1-x}$V$_{x}$, x=0.00, 0.02, 0.037.
Arrows indicate the Neel temperature. Inset shows $\emph{T}$$_{N}$
for Cr$_{1-x}$V$_{x}$ alloys, determined from the electrical
resistivity measurements reported in ~\cite{takeuchi1980} $\circ$,
and in the current studies $\bullet$.}
\end{figure}

While the minimum in the temperature dependence of $\rho$ in pure Cr
occurs at $\emph{T}$$_{N}$, neutron scattering experiments have not
yet confirmed the absence of the antiferromagnetism in
Cr$_{0.963}$V$_{0.037}$. Fig.~2a shows elastic longitudinal scans
along (100) direction in Cr$_{0.963}$V$_{0.037}$ and in pure Cr
collected at 5 K. The striking feature of Fig.~2a is the strong
scattering at the commensurate wavevector (100), absent in pure Cr.
Incommensurate scattering in Cr$_{0.963}$V$_{0.037}$ occurs at four
wavevectors q=(1$\pm$$\delta$$_{1,2}$00)2$\pi$/$\emph{a}$, where
$\delta$$_{1}$=0.078, $\delta$$_{2}$=0.058, which are all larger
than the wavevectors corresponding to antiferromagnetic ordering in
pure Cr. Both commensurate and incommensurate scattering in
Cr$_{0.963}$V$_{0.037}$ are resolution limited, which corresponds to
long range order in the crystal. Fig.~2b shows the temperature
dependence of the commensurate and incommensurate scattering in
Cr$_{0.963}$V$_{0.037}$ and the incommensurate scattering in pure
Cr. The commensurate scattering in Cr$_{0.963}$V$_{0.037}$ increases
monotonically upon cooling to 5 K. The incommensurate scattering
increases on cooling to $\sim$120 K, where it decreases and remains
finite upon further cooling to 5 K. The temperature dependence of
the incommensurate scattering in pure Cr is well understood. Below
311 K the spin density wave (SDW) is transversely polarized and the
scattering increases on cooling. At $\sim$120 K, the SDW changes
polarization to longitudinal, which results in the loss of the
component of the magnetic moment perpendicular to (100) in a single
magnetic domain sample. We observed a qualitatively similar
temperature dependence of incommensurate scattering in
Cr$_{0.963}$V$_{0.037}$, though the scattering remains finite below
$\sim$120 K. We conclude that contrary to the results of the
transport measurements, Cr$_{0.963}$V$_{0.037}$ orders
antiferromagnetically between 250 K and 300 K. Our findings indicate
that the temperature and wavevector dependence of elastic scattering
in Cr$_{0.963}$V$_{0.037}$ is very different from the one in pure Cr
and is likely to originate from the different nesting between the
hole and electron octahedra of the Fermi surface of
Cr$_{0.963}$V$_{0.037}$. Such mechanism was suggested previously to
account for observed commensurate and incommensurate scattering in
Cr$_{1-x}$Mn$_{x}$ alloys~\cite{sternlieb1994}.

\begin{figure}
\includegraphics[scale=0.42]{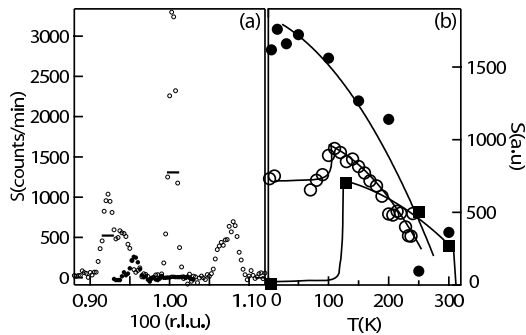}
\caption{\label{fig:epsart} (a) Elastic longitudinal scans [(100)
direction] in Cr$_{0.963}$V$_{0.037}$ ($\circ$) and in pure Cr
($\bullet$) at 5 K. Solid horizontal lines indicate spectrometer
resolution at the corresponding wavevectors. (b) Temperature
dependence of commensurate elastic scattering in
Cr$_{0.963}$V$_{0.037}$ ($\bullet$), incommensurate scattering in
Cr$_{0.963}$V$_{0.037}$ ($\circ$), and incommensurate scattering in
pure Cr ($\blacksquare$). Solid lines are a guide for the eye.}
\end{figure}

We have found strong temperature dependent elastic scattering at the
commensurate and incommensurate wavevectors in nominally quantum
critical Cr$_{0.963}$V$_{0.037}$. Although the results of transport
measurements reported in this work and other measurements reported
elsewhere indicate that Cr$_{0.963}$V$_{0.037}$ is a quantum
critical antiferromagnet, the neutron scattering unambiguously
establishes that the composition orders antiferromagnetically at
high temperature (T$>$250 K). The main effect of V doping on Cr is
not the reduction of the Neel temperature but rather the change in
nesting conditions of the Fermi surface and, perhaps the
restructuring of the Fermi surface itself.

Work at the Brookhaven National Laboratory is supported by the
Department of Energy and by the State University of New York, Stony
Brook. Work at the University of Michigan is supported by the
National Science Foundation under grant No. NSF-DMR-0405961. D. A.
S. and M. C. A. would like to thank A. M. Tsvelik and S. M. Shapiro
for useful discussions.


\begin{thebibliography}{99}

\bibitem{fawcett1988}E. Fawcett, Rev. Mod. Phys. $\bf{60}$ (1988) 209.
\bibitem{takeuchi1980}J. Takeuchi, H. Sasakura and Y. Masuda, J. Phys. Soc. Jpn $\bf{49}$ (1980) 508.
\bibitem{yeh2002}A. Yeh, Y. Soh, J. Brooke, G. Aeppli, T. F. Rosenbaum and S. M. Hayden, Nature, $\bf{419}$ (2002) 459.
\bibitem{lee2004}M. Lee, A. Husmann, T. F. Rosenbaum, and G. Aeppli, Phys. Rev. Lett., $\bf{92}$ (2004) 187201.
\bibitem{norman2003}M. R. Norman, Q. Si, Y. B. Bazaliy, and R. Ramazashvili, Phys. Rev. Lett., $\bf{90}$ (2003) 116601.
\bibitem{sternlieb1994}B. J. Sternlieb, E. Lorenzo, G. Shirane, S. A. Werner, E. Fawcett, Phys. Rev. B, $\bf{50}$ (1994) 16438.
Identification of commercial equipment in the text is not intended
to imply recommendation or endorsement by the National Institute of
Standards and Technology.
\end{thebibliography}
\end{document}